\documentstyle[11pt]{article}
\def\rmd{{\rm d}}
\def\etal{{\it et al.}}

\def\Asym{\mbox{{$A$\raisebox{-0.25em}{{\small{\small{\sl 0}}{\sl n}}}}}}

\def\degree{\mbox{$^\circ$}}

\def\dsdo{{\rmd \sigma\!/\!\rmd {\mit \Omega}}}

\newcommand{\pbarppipi}{\overline{p} p  \rightarrow \pi^- \pi^+}
\newcommand{\pbarptwopi}{\overline{p} p  \rightarrow \pi \pi}

\newcommand{\piNpiN}{\pi N \rightarrow \pi N}

\newcommand{\siml}{\parbox{0.8em}{\raisebox{0.3ex}
	{$<$}\hspace{-0.8em}\raisebox{-0.3em}{$\sim$}}}
\begin{document}

\begin{center}


\title
{Exclusive Hadronic Reactions at High $Q^2$ ($90\degree$)
     and Polarization Phenomena,
      An Experimental Proposal}

\vspace{5mm}

\author{
F. Myhrer \\
Department of Physics and Astronomy,
 University of South Carolina, \\
Columbia, SC29208  }

\date{}

\maketitle


{\bf Abstract.}

\end{center}

Arguments are presented for the expected behaviour of $\piNpiN$
scattering and the $\pbarppipi$ reaction at high energy and large
scattering angles.
The annihilation reaction has close to maximal asymmetry
($\approx$ 1) for $p_{lab} \siml $ 2.2 GeV/c.
As will be presented for fixed (90$\degree$) angle
this large asymmetry will
not become zero but will
start to oscillate
with energy at higher energies and large $Q^2$
when perturbative QCD becomes applicable.
This is due to the energy dependence of a QCD phase difference between
the independent quark-quark scattering (Landshoff) and short-distance
processes at high but not asymptotic energies.
A consequence of the existence the Landshoff process is that even
if helicity is conserved at the quark level ($m_{q}$ = 0 MeV),
helicity does not have to be conserved on the hadronic level.
We will discuss the implications  for spin observables
in $pp$ elastic scattering and argue that
these QCD phenomena are easier to explore theoretically in the
$\piNpiN$ scattering and/or in the crossed channel reaction
$\pbarppipi$ where the  analysis is simpler because
these two processes have only two helicity amplitudes.

\section{ Introduction}

For short-distance perturbative QCD exclusive hadronic
scattering processes the quarks are all connected by high $Q^2$
gluons and all quark propagators are far off-shell \cite{Brodsky}.

            \vspace*{6cm}
\noindent
\begin{description}
\item[{\rm Figure 1:}]
Illustrations of contributions to $\piNpiN$ scattering amplitudes.
On the left an example of
a short-distance contribution to $f_{SD}$.
On the right a contribution to the Landshoff amplitude $f_L$.
The dashed lines signify that this quark-quark scattering
does not have to occur in the same plane as the other scattering.
\end{description}

%
This means the short distance amplitudes $ f_{SD}$ are all real and no
polarization effects are expected.
However, the Landshoff amplitudes $ f_L$ will contribute
to the same exclusive hadronic scattering processes \cite{Landshoff}.
In the $ f_L$ amplitudes the hard gluons are also at high $Q^2$
but the two independent quark-quark scatterings
can take place in two parallel scattering planes
leading to the same final hadrons.
The distance between these two independent
quark-quark scatterings are determined by
the sizes of the hadrons involved.
The only requirement is that after the hard (high $Q^2$) scattering
the final quarks (antiquarks) move parallel with
roughly the same
speed to be able to form the final hadrons as
illustrated for $\piNpiN$ scattering in Fig.1.
The distance between the two quark-quark scatterings
implies one has a relative
angular momentum which can couple to the spin and give
for the hadronic reaction
at least a $L \cdot S$ amplitude.
Such an amplitude
violates helicity conservation on the hadronic level \cite{Ral92}.
Or said differently, since
the Landshoff amplitudes contain soft QCD processes
where a propagator is (almost) on-shell
(Sudakov form factors),
called "the Landshoff pinch" in the review by Mueller \cite{Mueller},
the amplitudes $ f_L$ will in general be complex.
As a consequence we will observe
polarization phenomena in hadronic reactions.


\section{ Asymmetries}

\subsection{ The Reaction $\pbarptwopi$}


        \vspace{45mm}
\noindent
\begin{description}
\item[{\rm Figure  2:}]
(a) An example of short-distance QCD diagram to
order $\alpha_s^4$ for the process $\pbarptwopi$.
The diagram has an $s^{-3}$ dependence.
(b) An example of diagrams for large-angle Landshoff process for the
same reaction of order $\alpha_s^3$.
The timelike gluon and one quark are off-shell and the diagram
gives an $s^{-5/2}$ behavior when we neglect radiative corrections.
\end{description}

First let us concentrate on the $\pbarppipi$ reaction
which has a large analysing power, $\Asym \approx $ 1 for
$p_{lab} \siml$ 2.2 GeV/c \cite{Ei75,Te89,Hasan92}
as discussed at this workshop \cite{Myh,TMK}.
If helicity is conserved on the hadronic level at very high energy,
then $\Asym$ should be zero at these energies.
This is correct {\it only if} the short distance amplitude $f_{SD}$
illustrated in Fig. 2a acts alone.
However,
as discussed in Ref. \cite{CCM},
the Landshoff amplitude, $f_{L}$ illustrated in Fig. 2b,
will contribute as well. Including the radiative corrections
$f_L$ will be at least of order $\alpha_s^4$ like
$f_{SD}$ illustrated in Fig. 2a,
but $f_L$  will fall off with increasing energy like $s^{2.85}$,
i.e., slower than $f_{SD}$.

The elementary quark-quark scattering amplitude has
an energy-dependent  phase, as
inferred by Ralston and Pire \cite{Ral82}
and calculated in perturbative QCD by Sen \cite{Sen}.
Its analytic form is
\begin{equation}
   \Phi \sim \frac{\pi}{6} \: ln \: ln \: (Q^2/ \Lambda^2)
\label{a}
\end{equation}
where $\Lambda \approx$ 100 MeV.
Ralston and Pire used this  phase in their
phenomenological hadronic Landshoff amplitudes
to describe the energy oscillations of the scaled $pp$ elastic
$90 \degree$ cross section.
They needed a constant $a$ ($ \approx 50$)
in front of the double log instead of $\pi/6$
of eq.(\ref{a}) to reproduce the observed (see Fig. 3)
period of the energy oscillations in the scaled $pp$
elastic 90$\degree$ scattering.
Botts and Sterman analysed this phase factor in hadronic reactions
and found the expression \cite{BS}
\begin{equation}
   \Phi = a \: ln \left( \frac{ln \: s/ \Lambda^2}{ln \:
           1/ (b\: \Lambda )^2} \right)  \: + \: {\rm constant} ,
\label{b}
\vspace{-2mm}
\end{equation}
where the constant a in perturbative QCD is $\pi /6$, and $\Lambda$
= 100 MeV as before.
The impact parameter $b$ can be thought of as the average
distance between the independent quark-quark scatterings. It has
the following energy dependence \cite{BS}
\begin{equation}
    b \: \Lambda = \: (\sqrt{s} \:/\Lambda^{\prime})^{ - \tau}
\label{c}
\vspace{-2mm}
\end{equation}
where $\tau \approx$ 0.7 for three flavors of quarks.
As discussed by Botts and Sterman \cite{BS,Botts}
the phase eq.(\ref{b}) should become independent of energy
at asymptotic energies ($s \rightarrow \infty$).

        \vspace{5cm}
\begin{description}
\item[{\rm Figure  3:}]
The elastic $pp$ cross section
at 90$\degree$ scaled by $s^{10}$ as a
function of energy. Figure taken from Carlson \etal\ \cite{CCM}.
\end{description}

\subsection{Elastic $pp$ Scattering}

If we apply these ideas to $pp$ elastic scattering
we can show that not only the oscillations in the
scaled cross section at $90\degree$, see Fig. 3,
can be reproduced (see here Ref. \cite{Ral82}), but also the
spin-correlation observable $A_{nn}$ at $90\degree$ can be described
\cite{CCM}.
The phenomenological arguments leading to these results are
following Ref. \cite{CCM}:
For elastic $pp$ scattering the
five helicity amplitudes  $M_i$ (i = 1, $\dots$ , 5) are
of the form
(the energy scale is factored out in $\phi_i$):
\begin{equation}
  \phi_i \propto s^{-4} \: M_i \: = \:
   s^{-4} \: ( B_i \: + \: C_i \: s^{0.2} \:
   e^{i[\Psi_i \: + \: \delta_i]} ),
\vspace{-2mm}
\end{equation}
where $B_i$, which originates from the short distance $pp$ amplitude
$M_{SD}$,
$C_i$ from the Landshoff amplitude $M_L$,
and $\delta_i$ are real constants.
The phase is deduced from eqs.(\ref{b}) and (\ref{c}) to be
\begin{equation}
  \Psi_i \: = \: a \: ln \left( \frac{ln(s/\Lambda^2)}
              {ln(s/\Lambda^2_i)} \right) .
\label{d}
\vspace{-2mm}
\end{equation}
The energy dependence of $A_{nn}$
at 90$\degree$ is then understood to be a "beating" of the different
energy-periods in the phases of the helicity amplitudes above \cite{CCM}.
With the interplay of the two amplitudes, $M_{SD}$ and $M_L$, it is not
difficult to reproduce the spin observables in $pp$ elastic
scattering. However, since $pp$ elastic scattering has
in general {\it five}
helicity amplitudes, there is too much freedom in fitting data.
Only for 90$\degree$ c.m. scattering do the expressions
simplify since $\phi_5$ = 0 and $\phi_4$ = - $\phi_3$ so we have only
three independent helicity amplitudes.

\section{Ideas for experimental proposals}

To examine if this phenomenological analysis
is reasonble, it would be preferable to
test the predictions in the two reactions $\piNpiN$ and
$\pbarptwopi$.
Each of these reactions are described by {\it only two} helicity
amplitudes, the helicity non-flip $f_{++}$ and
the helicity flip $f_{+-}$ amplitude.
Furthermore, these
measurements can be done at existing facilities
like $AGS$ at Brookhaven National Laboratory
or at Fermilab, and certainly
at the proposed facilities like $SuperLEAR$ or $KAON$.

\medskip

In terms of
these two helicity amplitudes
the cross section and the asymmetry are given as
\begin{equation}
   \dsdo =  | f_{++} |^2 +  |f_{+-} |^2
\hspace{0.8cm}{\rm and}\hspace{0.8cm}
\Asym = 2 \Im m(f_{++}^* f_{+-}) / (\dsdo).
\label{main}
\end{equation}
For the $\pbarptwopi$ reaction
the short-distance real amplitude $f_{SD}$ contributes only to
$f_{+-}$, whereas the Landshoff
amplitude $f_L$ contributes to both helicity amplitudes.
The energy dependences of the two amplitudes
are as follows:
\begin{equation}
   f_{SD} \propto s^{-3} \; \; \; {\rm  and} \;
     \; \; f_L \propto s^{-2.85}
\end{equation}
meaning the Landshoff amplitude will
also dominate at high enough energies for these reactions.

\medskip

For both reactions some
data already exist for the scaled cross sections
 $  s^8 \: \dsdo$
at $90\degree$.
For the reaction $\pbarppipi$ data exist for momenta
up to $p_{lab}$ =6.2 GeV/c \cite{Oslo1,Oslo2} as shown in Fig. 4
taken from Ref. \cite{CCM}.
The elastic $\piNpiN$ scattering
at $90\degree$ have been measured for momenta
as high as 30 GeV/c \cite{Oslo}.
In Fig. 5 we show the scaled cross section data
for the $\piNpiN$ scattering, a figure
taken from G. Blazey's thesis
\cite{Blazey}. Unfortunately the
highest energy measurement at $p_{lab}$ = 30 GeV/c has
uncertainties too large
to be useful in this discussion and is not shown in this figure.
As is clear from Figs. 4 and 5,
a few measurements at different energies with reasonable statistics
are needed to establish the possible oscillatory pattern of the
scaled cross section.

        \vspace{7cm}
\noindent
\begin{description}
\item[{\rm Figure  4:}]
The cross section for $\pbarppipi$ at 90$\degree$
scaled by $s^8$ as a function of $ln$ $s$. Figure taken from
Ref. \cite{CCM}.
\end{description}

\newpage

\vspace*{10cm}

\noindent
\begin{description}
\item[{\rm Figure  5:}]
The cross section for elastic $\piNpiN$ scattering at 90$\degree$
scaled by a factor $s^8$.
Figure taken from Ref. \cite{Blazey}.
\end{description}
\vspace*{3mm}

The question being asked is
if both of these scaled cross sections oscillate with energy
similar to what is observed for $pp$ elastic scattering, see Fig.3.
If this is found to be the case then
a further confirmation of the ideas presented here would be to
see similar energy oscillations in $A_{0n}$ for the same two
reactions.
Experimentally, the annihilation reaction might be
better since the
asymmetry at low energies
$p_{lab} \approx$ 2 GeV/c is very large
\cite{Ei75,Te89,Hasan92}.
However,
we do expect the geometric hadronic impact parameter
ideas used to explain this large asymmetry
\cite{Myh,TMK}
to break down
when the perturbative QCD regime of exclusive
hadronic reactions is reached at higher energies \cite{CCM}.
The onset of the perturbative QCD regime
may be signaled by
a significant change in the energy and
angular variation of the asymmetry, for example,
the large $\Asym$ at 90$\degree$ will become smaller and
start to oscillate with increasing energy if
the QCD phenomenology outlined above is reasonable.

\medskip

This work is supported in part by NSF grant no. PHYS-9006844.



\end{document}